\documentstyle[a4,12pt,epsf,rotate]{article}
\input{rotate}
\begin{document}
\begin{center}
\Large
{\bf The Triaxial Rotation Vibration Model in the Xe-Ba Region}\\[6pt]
\normalsize
\end{center}
\begin{center}
U. Meyer\footnote{Supported by the DFG Graduiertenkolleg MU 705/3}, Amand Faessler, S.B. Khadkikar\\[12pt]
{\it Institute for Theoretical Physics, University of T\"{u}bingen\\
Auf der Morgenstelle 14, D--72076 T\"{u}bingen, Germany}\\[10pt]
\end{center}
\begin{center}
{\bf Abstract}\\[10pt]
\end{center}
\begin{center}
\begin{minipage}[b]{10.5cm}
\small
{Collective quadrupole degrees of freedom give rise to vibrations and rotations
in nuclei. The axial Rotation Vibration Model (RVM) is here extended to 
describe also triaxial equilibrium shapes with $\beta$ and $\gamma$ 
vibrations allowing for the interaction between vibrations and rotations. This
Triaxial Rotation Vibration Model (TRVM) is applied to Ba and Xe isotopes
with A$\approx$120 to 130. This area has recently been pointed out 
to be a region
for the O(6) limit of the Interacting Boson Approximation (IBA). The present
work shows that the TRVM can equally well describe these nuclei concerning
their excitation energies and E2 branching ratios.}\\[24pt]
\end{minipage}
\end{center}
\newpage
\section{Introduction}
The collective quadrupole degrees in nuclei yield rotational and vibrational 
excitations. They are described by the Bohr-Mottelson Hamiltonian \cite{BM}.
 The wavefunctions for rotations and $\beta$ and $\gamma$ vibrations in
strongly deformed axially symmetric nuclei have been greatly improved in the
Rotation Vibration Model (RVM) by Faessler and Greiner \cite{Fae1,Fae2,Fae3,Fae4} taking the
interaction between the rotations and vibrations into account with an axially
symmetric equilibrium deformation. \par
Recently, the Ba and Xe region with mass numbers A$\approx$120 to 130 has been
studied experimentally [6, 9-15] and interpreted by several 
models: They compare 
different approaches of the Interacting Boson Approximation (IBA) in the
U(5) and the O(6) limits (with different E2 operators) \cite{Iachello} 
and the axial Rotation
Vibration Model (RVM) and the Asymmetric Rotor Model (ARM) \cite{Dawydow}. They
exclude the ARM \cite{Lie,Sei} because it cannot describe a K=0 band
built on $\gamma$ vibrations \cite{Sei} and because of the wrong staggering
of the $2^{+}$, $3^{+}$, $4^{+}$, $5^{+}$, $\ldots$ excitation energies
in the K=2 band. The argument with the K=0 band built on $\gamma$ vibrations
is trivial: The ARM does not contain $\gamma$ vibrations and thus cannot
describe it. Because the staggering in the ($K$=2) 
quasi $\gamma$ band is not
described correctly in the ARM, but given correctly in the RVM, the
staggering seems to be due to rotation vibration
interaction.
 Although the RVM \cite{Fae1,Fae2,Fae3,Fae4} does quite well in describing the data, the Cologne
group concludes \cite{Sei} that the O(6) IBA with the additional parameter
$\chi$ in the E2 transition operator does in average agree better with the
data. Here, we want to show that this is connected with the restriction
of the RVM \cite{Fae1,Fae2,Fae3,Fae4} to axial symmetry while IBA allows also for triaxiallity
(O(6) = $\gamma$ instable limit). We present here an extension of the
axial RVM to triaxiallity (Triaxial Rotation Vibration Model = TRVM). This
model has the same number of parameters as the IBA used \cite{Lie,Sei}. We obtain
an equally good agreement with the data as in IBA.\par
In Section 2 we present the model which is an extension of the RVM
\cite{Fae1,Fae2,Fae3,Fae4} 
allowing for a triaxial equilibrium deformation (TRVM). In Section 3 we compare
the excitation energies and the branching ratios in the best measured
Xe and Ba isotopes. Section 4 summarizes the main results. 
\section{The Triaxial Rotation Vibration Model}
In the Rotation
Vibration Model \cite{Fae1,Fae2,Fae3,Fae4} one characterizes as in the Bohr-Mottelson Model
\cite{BM} the surface of the nucleus by quadrupole deformations:
\begin{equation}
R(\theta,\phi)=R_0(1+\sum_{\mu}\alpha_{2\mu}Y_{2\mu}(\theta,\phi))
\end{equation}
The deformation parameters $\alpha_{2\mu}$ are considered as dynamical 
variables and depend classically on time. Their behaviour is governed
by the Hamiltonian:
\begin{equation}\label{Ham}
H=\frac{1}{2}B\sum_{\mu}\dot{\alpha}_{2\mu}^{\dag}\dot{\alpha}_{2\mu}+V(\alpha_{2\mu})
\end{equation}
The information about the equilibrium shape of the nucleus is contained
in the potential energy $V(\alpha_{2\mu})$. The five quadrupole degrees of 
freedom $\alpha_{2\mu}$ can be replaced by the three Euler angles 
$(\phi, \theta, \psi)$ for rotations and two deformation parameters
\begin{equation}\label{trans}
\alpha_{2\mu}={\cal D}_{\mu 0}^2(\phi, \theta, \psi) (\beta_0 + a_0^{\prime}(t))+
({\cal D}_{\mu 2}^2+{\cal D}_{\mu -2}^2) (a_2 + a_2^{\prime}(t))\hspace*{1cm}.
\end{equation}
Here $\beta_0$ and $a_2$ give the equilibrium shape of the nucleus while
$a_0^{\prime}(t)$ and $a_2^{\prime}(t)$ describe vibrations around this shape.
 They are connected with the Bohr-Mottelson parameters through:
\begin{eqnarray}
\beta_0 + a_0^{\prime}(t) & = & \beta cos(\gamma)\nonumber\\
a_2 + a_2^{\prime}(t) & = & \frac{1}{\sqrt{2}}\beta sin(\gamma)
\end{eqnarray}
In deriving the Hamiltonian for the TRVM we follow very closely the
derivation for the axial RVM in Ref.\cite{Fae3}. With the 
transformation (\ref{trans})
the Hamiltonian (\ref{Ham}) can be written (see in Ref.\cite{Fae3} 
 eqs. (3) and (4)):
\begin{equation}\label{Ham1}
H=T+V=T_{rot}+T_{vib}+T_{rotvib}+V_{a_0 a_2}(a_0^{\prime}, a_2^{\prime})
\end{equation}
The different terms are obtained by the straight forward transformation 
(\ref{trans}) and the expansion in powers of $a_0^{\prime}/b_0$ and
$a_2^{\prime}/a_2$ up to second order. This assumes that 
the $\beta$ and $\gamma$ deformation is large compared to the 
vibrational amplitudes.
This approach therefore does not allow to describe spherical or nearly
spherical nuclei. The several terms in the TRVM Hamiltonian are
given by:
\begin{eqnarray}\label{Ham2}
T_{rot} & = & \frac{\hat{{\bf I}}^2-\hat{I}_3^2}{2I_0}+
\frac{\hat{I}_3^2}{16Ba_2^2}\nonumber\\
T_{vib} & = & -\frac{\hbar^2}{2B}(\frac{\partial^2}{\partial a_0^{\prime 2}}+
\frac{1}{2}\frac{\partial^2}{\partial a_2^{\prime 2}})\nonumber\\
T_{rotvib} & = &   \frac{\hat{{\bf I}}^2-\hat{I_3}^2}{2I_0}f_0(\beta_0,a_2,a_0^
{\prime},a_2^{\prime})\\
{} & {} &
+ \frac{\hat{I}_{+}^2+\hat{I}_{-}^2}{2I_0}f_1(\beta_0,a_2,a_0^
{\prime},a_2^{\prime})\nonumber\\
{} & {} &
+ \frac{\hat{I}_3^2}{16Ba_2^2}f_2(a_2,a_2^{\prime})+2\epsilon\frac{a_0^{\prime}}{\beta_0}
\nonumber
\end{eqnarray}
The functions $f_0$, $f_1$, and $f_2$ are  obtained by the expansion mentioned
above:
\begin{eqnarray}
f_0 & = & -2\frac{a_0^{\prime}}{\beta_0}+3\frac{a_0^{\prime^{2}}}{\beta_0^2}+
\frac{2}{\beta_0^2}(a_2^2+2a_2a_2^{\prime}+a_2^{\prime^2})
\nonumber\\
f_1 & = & \frac{1}{3}\sqrt{6}\frac{1}{\beta_0}(a_2+a_2^{\prime})-
\sqrt{6}\frac{1}
{\beta_0^2}a_0^{\prime}(a_2+a_2^{\prime})\nonumber
\\
f_2 & = & -2\frac{a_2^{\prime}}{a_2}+3\frac{a_2^{\prime^2}}{a_2^2}\nonumber
\end{eqnarray}
For the potential energy we assume a harmonic oscillator potential
around the equilibrium shape:
\begin{equation}\label{Ham3}
V(a_0^{\prime},a_2^{\prime})=\frac{1}{2}C_0a_0^{\prime^2}+C_2a_2^{\prime^2}
\end{equation}
For the diagonalization of the Hamiltonian we 
chose a basis of eigenstates of the Hamiltonian $H_0=T_{rot}
+T_{vib}+V$.  
The eigenstates of the unperturbed Hamiltonian $H_0$ are easily obtained
and labelled by the total angular momentum $I$, the projection on the
intrinsic 3-axis $K$ and the number of phonons for the $\beta$-$(n_0)$ and
$\gamma$ vibrations $(n_2)$:
\begin{equation}\label{basis}
|IK,n_2 n_0\rangle= \left(\frac{2I+1}{16\pi^2}\frac{1}{1+\delta_{K0}}
\right)^{\frac{1}{2}}({\cal D}_{MK}^I+(-)^I{\cal D}_{M-K}^I)
\sqrt{\frac{1}{n_0 !}}\left(\hat{\beta}_0^{\dag}\right)^{n_0}
|0\rangle 
\sqrt{\frac{1}{n_2 !}}\left(\hat{\beta}_2^{\dag}\right)^{n_2}
|0\rangle 
\end{equation}
Due to the rotation vibration part of the Hamiltonian $T_{rotvib}$ 
several eigenstates of $H_0$ are mixed. 
 The unperturbed energies as eigenvalues of $H_0$ 
may easily be obtained. They are given by:
\begin{eqnarray}
E^{IK}_{n_2n_0} & = & (I(I+1)-K^2)\frac{\hbar^2}{2I_0}+\frac{K^2}{16Ba_2^2}+
(n_0+\frac{1}{2})E_{\beta}+(n_2+\frac{1}{2})E_{\gamma}\nonumber\\
\mbox{with}\hspace*{0.5cm}I& = & \left\{\begin{array}{lc}
0,2,4,6,\dots\ldots\ldots\ldots & \mbox{for K=0}\\
K,K+1,K+2,\ldots & \mbox{for K$\neq$0}
\end{array}\right.\nonumber\\
E_{\beta}=\hbar\sqrt{\frac{C_0}{B}}& ; &E_{\gamma}=\hbar\sqrt{\frac{C_2}
{B}}\\
\epsilon& = &\frac{\hbar^2}{I_0}\nonumber\\
I_0 & = & 3B\beta_0^2\nonumber
\end{eqnarray}
The operators $\hat{\beta}^{\dag}$ are creation 
operators for harmonic oscillations. 
We diagonalize the Hamiltonian (\ref{Ham1}), (\ref{Ham2}), (\ref{Ham3}) in
the complete basis (\ref{basis}).
 For numerical reasons the Hilbert space has to be truncated. We chose the
31 lowest basis states with $K\leq 6$ and up to two phonons
($n_2+n_0\leq 2$). In notation (\ref{basis}) the most important
states $|IK,n_2 n_0\rangle$ are:
\begin{eqnarray}
|I0,00\rangle & \ldots\ldots & \mbox{ground state band}\nonumber\\
|I2,00\rangle & \ldots\ldots & \mbox{K=2 quasi}\hspace*{1mm}\gamma
\hspace*{1mm}\mbox{band}\nonumber\\
|I0,10\rangle & \ldots\ldots & \mbox{one-phonon}\hspace*{1mm}\gamma
\hspace*{1mm}\mbox{band}\\
|I0,01\rangle & \ldots\ldots & \mbox{one-phonon}\hspace*{1mm}\beta
\hspace*{1mm}\mbox{band}\nonumber\\
|I4,00\rangle & \ldots\ldots & \mbox{K=4 quasi}\hspace*{1mm}\gamma
\hspace*{1mm}\mbox{band}\nonumber
\end{eqnarray}
 The TRVM has as the IBA four
independent parameters. 
 The vibration energies $E_{\beta}$ and $E_{\gamma}$, the inverse
moment of inertia $\epsilon=1/I_0$ and the triaxial equilibrium deformation
$a_2$. \par
The Hilbert space is too limited to describe the full variation of the 
moment of inertia. As in the competing IBA \cite{Kir} we make for the
energies the Lipas ansatz 
\begin{equation}
E=E_0/(1+\alpha_L E_0)\hspace*{1cm}
\end{equation}
$E_0$ is the excitation energy obtained after diagonalization. The Lipas
parameter $\alpha_L$ is quite small ($\approx 10^{-4}$[keV$^{-1}$]). It 
describes a variable moment of
inertia. We would like to note that the number of parameters including the
Lipas parameter is the same in TRVM and in IBA (including the effective
boson charge in IBA). In the O(6) limit the branching ratios of IBA depend 
only on the parameter $\chi$. But similarly most of the branching ratios 
in TRVM depend essentially only on the effective triaxiallity $a_2+a_2^{\prime}
$. In addition, one could omit practically the $\beta$ band and therefore the
parameter $E_{\beta}$ in TRVM, since the results for the branching ratios 
do almost not depend on
the $\beta$ band and the agreement would be equally good by omitting this
band. But on the other side, this means also that our prediction for the
$\beta$ bandhead is not reliable. Both 
models therefore effectively have the same number of 
parameters.
\section{Comparison with the Data}
\subsection{Energy spectra}
In Figure 1 we display our results for the energy spectrum of $^{130}$Ba.
 The free parameters which give an overall best fit of the experimental
data from Ref.\cite{Kir} are listed in Table 1. 
 For comparison, in Figure 1 furthermore the calculated IBA values 
and the experimental results are shown. For the low spin states 
in the ground state band both 
theoretical results are in nice agreement with experiment. As a result
of the applied Lipas fitting procedure lower spin states as the 4$^+$,
 6$^+$ and also the 8$^+$ state lie slightly higher while the 
10$^+$-state is little below the experimental values. This is due to
the stronger effect of the Lipas procedure on higher spin states.\par
For the excited $K$=2 band (quasi $\gamma$ band) we obtain in TRVM the
experimentally observed staggering of even-odd angular momentum states.
 A pure triaxial rotor model without vibrations 
but with a $\gamma$ deformation of the nuclear surface
does not exhibit this feature. The staggering thus seems to be the result
of the rotation vibration coupling. We would like to note that the staggering
increases with excitation energies in TRVM. In contrast, in IBA 
the staggering seems
to be constant with growing excitation energy. Both models 
agree qualitatively with the data in the $\gamma$ band but not in detail.\par
 The $K$=0 band is interpreted in the TRVM as the $\gamma$ one-phonon band.
 The bandhead ($0^+$) at 1179 keV is fitted through the vibration energy
$E_{\gamma}$. For the TRVM the 0$^+$--2$^+$ splitting is nicely reproduced
and in better agreement with experiment than IBA. The $\beta$ vibrational
bandhead is at around 1150 keV. Up to now, the levels of this band have 
not been
identified in the experimental analysis which might be difficult because
the two 0$^+$ energies are very close. Since our values for the experimentally
observed energy levels are almost not affected by the $\beta$ band, we do not
display it in Figure 1.\par
\marginpar{Table 1}
\marginpar{Figure1}
 In Figure 2 we show our results for the $^{126}$Xe spectrum. 
There a different set of parameters must be used in order to fit 
the experimental data best. This set is displayed in Table 2. The 
agreement
with the experimental level spacings is in the two theoretical models 
about the same. This holds for the ground state band as well as for 
the quasi $\gamma$ band.
 In TRVM the staggering in low lying states is rather low. The free parameters
$a_2 /\beta_0$ and $E_{\beta}$ have to be increased relative to these
in $^{130}$Ba to get the best agreement. The excited $0^+$ band is in
good agreement with experiment and the level spacings obtained in TRVM
are in better agreement with the experiment than the IBA
calculations.\\[12pt]
\marginpar{Table 2}
\marginpar{Figure 2}
\subsection{E2 Branching Ratios}
The electric quadrupole operator can again be obtained from the one
in Ref.\cite{Fae3} by replacing $a_2^{\prime}\rightarrow a_2+a_2^{\prime}$.
 In collective intrinsic coordinates, it is given by:
\begin{eqnarray}
m(E2,\mu)& = &\frac{3Z}{4\pi}R_0^2\left[{\cal D}_{\mu 0}^2\left(\beta_0
\left(1+\frac{2}{7}\left(\frac{5}{\pi}\right)^{\frac{1}{2}}\beta_0\right)
\right)+{\cal D}_{\mu 0}^2 a_0^{\prime}\left(1+\frac{4}{7}
\left(\frac{5}{\pi}\right)^{\frac{1}{2}}\beta_0\right)+\right.\nonumber\\
{}&{}& {\cal D}_{\mu 0}^2\frac{2}{7}\left(\frac{5}{\pi}\right)^{\frac{1}{2}}
(a_0{\prime^2}-2(a_2+a_2^{\prime})^2)+\\
{}&{}&\left.
({\cal D}_{\mu 2}^2+
{\cal D}_{\mu -2}^2)\left(\left(1-\frac{4}{7}\left(\frac{5}
{\pi}\right)^{\frac{1}{2}}\beta_0\right)(a_2+a_2^{\prime})-
\frac{4}{7}\left(\frac{5}{\pi}\right)^{\frac{1}{2}}a_0^{\prime}(a_2+
a_2^{\prime})\right)\right]\nonumber
\end{eqnarray}
The reduced E2 transition probabilities are defined by 
\cite{Fae1,Fae2,Fae3,Fae4}:
\begin{equation}
B(E2;I_i\rightarrow I_f)=\frac{2I_f+1}{2I_i+1}|
\langle I_i||m(E2)||I_f\rangle|^2
\end{equation}
  We have calculated the measured B(E2) ratios to be able to compare them
with the experimental results for both $^{130}$Ba and $^{126}$Xe. In Table
3 the results for $^{130}$Ba are given. We see a fair agreement between the
results of TRVM and the experiment as well as those of IBA except for
the transition from the initial $4^{+}_{3}$ state. Here our result is about two orders of
magnitude larger while the IBA result is two orders of magnitude lower
than the experimental ratio B(E2,$4^{+}_{3}\rightarrow 2^{+}_{2}$)/
B(E2,$4^{+}_{3}\rightarrow 4^{+}_{2}$). A similar
discrepancy exists also for the transition ($4^+_3 \rightarrow  2^+_1$) 
as can be seen from the
last entry in Table 3. Any B(M1) contribution to the $4^+_3\rightarrow 4^+_2$
transition cannot cure this disagreement.\marginpar{Table 3}\\[12pt]
 Table 4 shows the ratios of certain B(E2) transitions in $^{126}$Xe .
Here the agreement is much better between the models and the
experiment; and both Interacting Boson Approximation  and the
Triaxial Rotation Vibration Model agree nicely with the data.
This might be because only pure quadrupole transitions are used
in Xe to define the ratios, while the uncertainty in the E2/M1 ratio
in the normalising states is ignored in the case of Ba and transitions
are assumed to be pure E2 transitions.\\[12pt]
\marginpar{Table 4}
\section{Conclusions}
The Triaxial Rotation Vibration Model (TRVM) is successfully applied to
Ba and Xe nuclei to obtain low lying energy levels. These results compare
well with those obtained in the O(6)-limit of IBA and with the experimental
values both for the spectra and for relative E2 transition rates. The bandhead
of the $\beta$ vibrational band are not seen in experiment so far. 
 Unfortunately, we are not able to specify very exactly the energy 
region for the bandhead or other levels of the $\beta$ band
since our calculated values are not very sensitive to the
parameter $E_{\beta}$.\\[12pt]
{\bf Acknowledgment}: We would like to thank Dr.\ Ingo Wiedenh\"{o}ver and
Prof.\ Peter von Brentano for providing us the IBA results and for fruitful 
discussions.

\newpage

\begin{center}
  \leavevmode
  \epsfxsize=12cm
  \epsfysize=15cm
  \setbox0=\vbox{\hbox{
  \epsfbox[33 60 540 750]{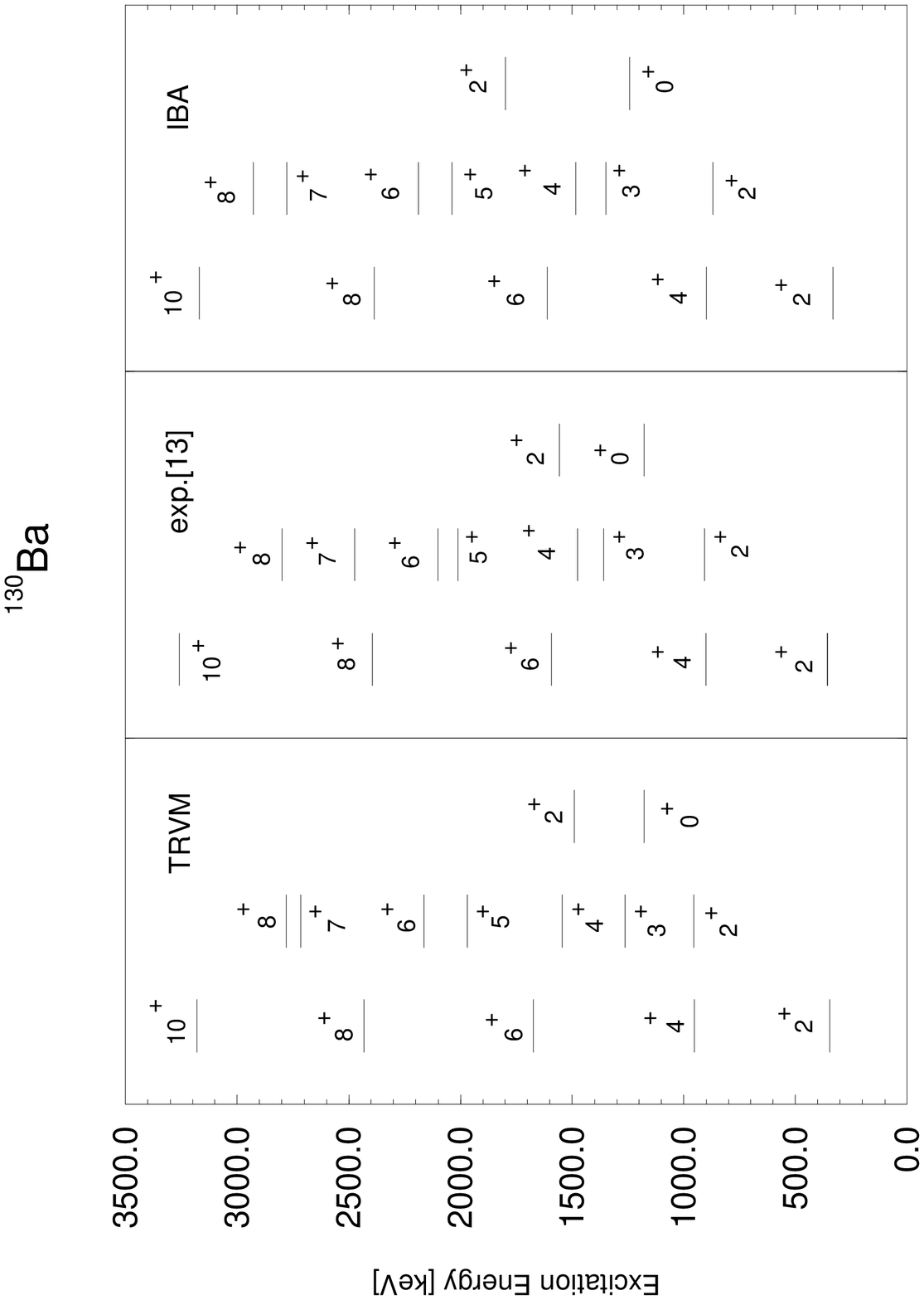}
}}
  \rotr0  
\end{center}
\vspace*{-1.0cm}
\begin{minipage}[b]{16.0cm}
\hspace*{2.45cm}
{\small Fig.1: 
Ground state, quasi $\gamma$ and $K$=0 band for $^{130}$Ba in keV for the 
\hspace*{2.55cm} IBA calculation and the
Triaxial Rotation Vibration Model (TRVM) compared \hspace*{2.60cm}
with experimental values from Ref.\cite{Kir}.}\\[24pt]
\end{minipage}

\newpage
\begin{center}
  \leavevmode
  \epsfxsize=12cm
  \epsfysize=15cm
  \setbox0=\vbox{\hbox{
  \epsfbox[33 60 540 750]{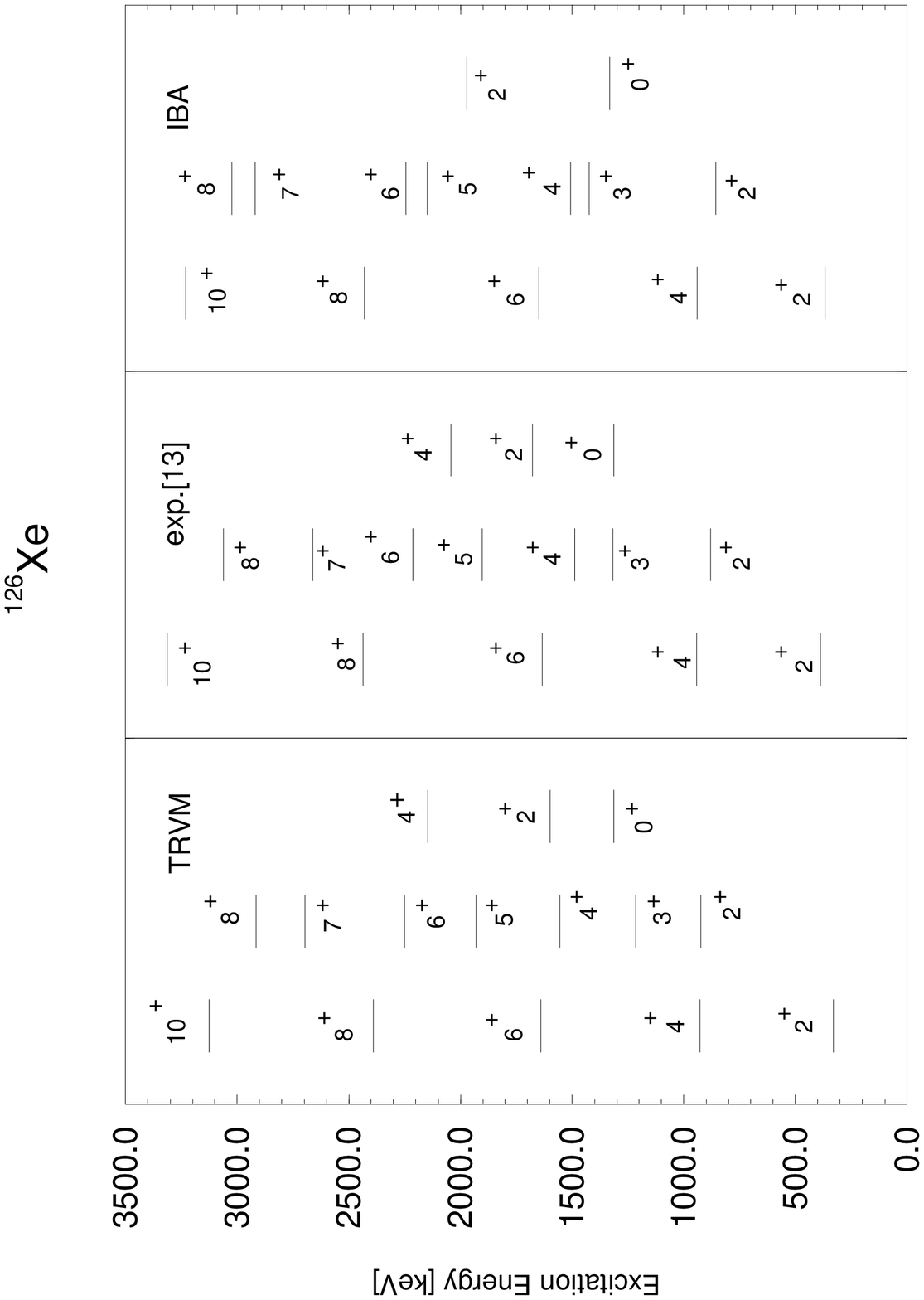}
}}
  \rotr0  
\end{center}
\vspace*{-1.0cm} 
\begin{minipage}[b]{16.0cm}
\hspace*{2.45cm}
{\small Fig.2: Ground state, quasi $\gamma$ and $K$=0 band for $^{126}$Xe 
in keV for the \hspace*{2.55cm}
IBA calculation and the
Triaxial Rotation Vibration Model (TRVM) compared \hspace*{2.60cm}
with experimental values from Ref.\cite{Kir}.}\\[24pt]
\end{minipage}\newline
\newpage
\vspace*{-1cm}
\begin{center}
\large
\begin{tabular}{|c|c|c|c|c|}
\hline
$E_{\beta}$[keV] & $E_{\gamma}$[keV] & $\epsilon$[keV] & $a_2/\beta_0$ & $\alpha_L$[keV$^{-1}$]\\
\hline
1168 & 1179 & 80.5 & 0.329 & $10^{-4}$\\
\hline
\end{tabular}\\[6pt]
\normalsize
\begin{minipage}[b]{10cm}
{\small Table 1: Parameters of the TRVM in the present 
calculations for $^{130}$Ba. $E_{\beta}$ and $E_{\gamma}$ are the vibration 
energies, $\epsilon$ is the inverse
moment of inertia and $a_2/\beta_0$ the relative ratio of the static
$\gamma$ and $\beta$
deformations. $\alpha_L$ is the Lipas parameter.}\\[6pt]
\end{minipage}\\[6pt]
\end{center}
\vspace*{-1cm}
\begin{center}
\large
\begin{tabular}{|c|c|c|c|c|}
\hline
$E_{\beta}$[keV] & $E_{\gamma}$[keV] & $\epsilon$[keV] & $a_2/\beta_0$ & $\alpha_L$[keV$^{-1}$]\\
\hline
1769 & 1314 & 81.3 & 0.333 & $10^{-4}$\\
\hline
\end{tabular}\\[6pt]
\normalsize
\begin{minipage}[b]{10cm}
{\small Table 2: Parameters of the TRVM in the present 
calculations for $^{126}$Xe. $E_{\beta}$ and $E_{\gamma}$ are the vibration 
energies, $\epsilon$ is the inverse
moment of inertia and $a_2/\beta_0$ the relative ratio of the static
$\gamma$ and $\beta$
deformations. $\alpha_L$ is the Lipas parameter.}\\[6pt]
\end{minipage}
\end{center}

\vspace*{-1cm}
\begin{center}
\begin{small}
\begin{tabular}{|r|r|r|r|}
\hline
{}&{}&{}&{}\\
$I_{i}\rightarrow I_{f}$&IBA--CQF&exp.\cite{Kir}&TRVM\\
{}&{}&{}&{}\\
\hline
$2_{2}^{+}\rightarrow 0_{1}^{+}$&6.2&$6.2\pm 0.7$&4.5\\
$2_{1}^{+}$&100.0&100.0&100.0\\
$3_{1}^{+}\rightarrow 2_{2}^{+}$&100.0&100.0&100.0\\
$4_{1}^{+}$&29&$22.0\pm 3.0$&16\\
$2_{1}^{+}$&6.1&$4.5\pm 0.6$&2.6\\
$4_{2}^{+}\rightarrow 2_{2}^{+}$&100.0&100.0&100.0\\
$4_{1}^{+}$&58&$54.0\pm 10.0$&57\\
$2_{1}^{+}$&0.12&$2.3\pm 0.4$&2.75\\
$0_{2}^{+}\rightarrow 2_{2}^{+}$&100.0&100.0&100.0\\
$2_{1}^{+}$&2.5&$3.3\pm 0.2$&3.0\\
$2_{3}^{+}\rightarrow 0_{2}^{+}$&100.0&100.0&100.0\\
$2_{2}^{+}$&0.11&$21.0\pm 4.0$&0.04\\
$4_{1}^{+}$&1.3&$2.7\pm 0.5$&0.11\\
$2_{1}^{+}$&0.0&$3.3\pm 0.6$&0.06\\
$0_{1}^{+}$&0.24&$0.017\pm 0.003$&0.04\\
$3_{1}^{+}$&112&${}$&18\\
$4_{3}^{+}\rightarrow 2_{2}^{+}$&0.02&$2.9(5)$&670\\
$3_{1}^{+}$&108&$97(17)$&214\\
$4_{2}^{+}$&100.0&$100.0^{*}$&100.0\\
$4_{1}^{+}$&0.01&$3.4(6)^{*}$&7.2\\
$2_{1}^{+}$&0.0&$0.30(6)$&23.3\\
\hline
\end{tabular}\\[6pt]
\end{small}
\begin{minipage}[b]{8cm}
{\small Table 3: B(E2) branching ratios of the TRVM for $^{130}$Ba in comparison with IBA and with experiment from Ref.\cite{Kir}.}
\end{minipage}
\end{center}

\begin{center}
\begin{tabular}{|r|r|r|r|}
\hline
{}&{}&{}&{}\\
$I_{i}\rightarrow I_{f}$&IBA O(6)$_{\chi,\gamma}$&{}exp.\cite{Sei}&{}{}{}{}TRVM\\
{}&{}&{}&{}\\
\hline
$2_{2}^{+}\rightarrow 0_{1}^{+}$&1.51&$1.5\pm 0.4$&8.7\\
$2_{1}^{+}$&100.0&100.0&100.0\\
$3_{1}^{+}\rightarrow 2_{2}^{+}$&100.0&100.0&100.0\\
$4_{1}^{+}$&41&$34.0^{+10}_{-34}$&24\\
$2_{1}^{+}$&1.9&$2.0^{+0.6}_{-1.7}$&4.4\\
$4_{2}^{+}\rightarrow 2_{2}^{+}$&100.0&100.0&100.0\\
$4_{1}^{+}$&89&$76.0\pm 22.0$&66\\
$2_{1}^{+}$&2.1&$0.4\pm 0.1$&1.5\\
$0_{2}^{+}\rightarrow 2_{2}^{+}$&100.0&100.0&100.0\\
$2_{1}^{+}$&2.4&$7.7\pm 2.2$&1.1\\
$2_{3}^{+}\rightarrow 0_{2}^{+}$&100.0&100.0&100.0\\
$2_{2}^{+}$&2.7&$2.2\pm 1.0^{*}$&0.8\\
$4_{1}^{+}$&0.95&$2.0\pm 0.8$&0.04\\
$2_{1}^{+}$&0.8&$0.14\pm 0.06^{*}$&0.01\\
$0_{1}^{+}$&0.01&$0.13\pm 0.04$&0.01\\
$3_{1}^{+}$&86&$67.0\pm 22.0^{*}$&20\\
$5_{1}^{+}\rightarrow 6_{1}^{+}$&49&$75\pm 23$&43\\
$4_{2}^{+}$&52&$76\pm 21$&83\\
$3_{1}^{+}$&100.0&100.0&100.0\\
$4_{1}^{+}$&2.3&$2.9 \pm 0.8$&0.8\\
$6_{2}^{+}\rightarrow 6_{1}^{+}$&43&$34^{+15}_{-25}$&22\\
$4_{2}^{+}$&100.0&100.0&100.0\\
$4_{1}^{+}$&1.0&$0.49\pm 0.15$&1.3\\
$7_{1}^{+}\rightarrow 6_{2}^{+}$&20&$40\pm 26$&22\\
$5_{1}^{+}$&100.0&100.0&100.0\\
\hline
\end{tabular}\\[6pt]
\begin{minipage}[b]{9cm}
{\small Table 4: B(E2) branching ratios of the TRVM for $^{126}$Xe in 
comparison with the IBA and with experiment from Ref.\cite{Sei}.}
\end{minipage}
\end{center}

\end{document}